\documentclass[reprint,amsmath,amssymb,aps,prl,floatfix]{revtex4-2}
\usepackage{float}
\usepackage[normalem]{ulem}
\usepackage{color}
\usepackage[section]{placeins}
\usepackage{graphicx}
\usepackage{dcolumn}
\usepackage{bm}
\graphicspath{ {./figures/} }

\begin{document}
	

	\title{Corrections to Conformal Anomaly at BKT Transitions}
	
	\author{Jon Spalding$^{1}$}
	\email{jondavid.spalding@ntu.edu.sg}
	

	\affiliation{$^1$School of Physical and Mathematical Sciences, Nanyang Technological University, Singapore}

	\date{\today}
	
	\begin{abstract}
		The conformal anomaly is key to describing the physics of interacting field theories in two dimensions, and has been shown to obtain finite-size corrections that are useful to numerical and experimental studies. At Berezinskii-Kosterlitz-Thouless transitions, the corrections are logarithmic with previously unknown universal coefficients. In this paper, I reveal the universal, $1^{st}$-order correction to the conformal anomaly for finite-sized systems at BKT transitions. This term results solely from topological defects that are irrelevant in the infrared limit. I compare the result with careful observations of entanglement for the quantum Heisenberg model with periodic boundary conditions in one spatial dimension. 
	\end{abstract}
	
	\pacs{Valid PACS appear here}
	\maketitle

	\textit{Introduction}
	One-dimensional quantum and two-dimensional classical critical theories exhibit conformal invariance, with many properties involving a universal quantity known as ``c" which is the central charge from the stress-energy tensor Virasoro algebra or conformal anomaly from the stress-energy tensor correlation function, respectively. It appears as the response of a theory to changes in metric, provides the first universal finite-size correction term to the free energy for classical systems, and the specific heat capacity for quantum systems. It can also be interpreted as the number of bosonic fields in a theory \cite{FMSCFT,PhysRevLett.56.742,PhysRevLett.56.746}.
	
	``c" gained attention as a coefficient on the universal logarithmic contribution to the bipartite von-Neumann entanglement entropy of ground-state (pure) wavefunctions in one spatial dimension \cite{HOLZHEY1994443,PhysRevLett.90.227902,PhysRevLett.92.096402} and renyi entropies \cite{jin2004quantum} and at finite temperatures \cite{korepin2004universality}. This paralleled the development of DMRG and MPS methods \cite{white1992density,white1993density,SCHOLLWOCK2011} which implicitly rely on entanglement entropy to minimize a variational wavefunction. These methods provide convenient access to high-precision measurements of entanglement with far greater ease than any other observable in an MPS or other numerical method such as QMC \cite{hastings2010measuring}. This feature motivates the numerical component of the current study.
	
	Gaussian critical points in one quantum or two classical dimensions formed the cornerstone of the renormalization group paradigm for both condensed and particle physics. But shortly afterwards, the new paradigm of topology took over, beginning with BKT transitions. These were the first topological phase transitions discovered for interacting many-body systems, rewarded by a Nobel prize in 2016
	\cite{kosterlitz2018ordering,berezinskii1971destruction,berezinskii1972destruction}. Since these initial theoretical discoveries, the field of topological physics in matter has continued growing.
	
	More recently, analytical calculations based on conformal field theory for Gaussian criticality provided the precise form of the entanglement entropy with open and periodic boundaries in infinite-size chains, before finite-size effects were observed numerically \cite{Laflorencie2006} and then treated analytically \cite{JStats2010, JStats2011P01017}. For a finite-size chain with periodic boundaries,

	\begin{equation}\label{eq:SemiInfiniteEntropy}
		S_{vN} = S_0 + \frac{c}{3}\ell\text{.}
	\end{equation}

	Here I define $\ell \equiv log(L/\pi sin(\pi x/L))$ as the length scale for periodic boundaries, and point out that the coefficient c simply indicates the relative concavity, or amplitude of the curvature, of a plot of entanglement. In \cite{JStats2010}, it was determined that in the case of a marginally irrelevant operator (the case at a BKT transition), the central charge obtains a correction proportional to $log(\ell)^{-3}$. This term can be re-interpreted as a finite-size correction to c itself in addition to a contribution to the entanglement entropy at BKT transitions as described in \cite{calabrese2009entanglement,casini2004,Laflorencie2006,PhysRevLett.104.095701}.
	
	If the exact form of this correction were known, including universal coefficients, it would provide useful information at classical BKT transitions in two dimensions as well as quantum BKT transitions in one dimension. In the quantum case, it would enable precise and convenient identification of BKT transitions from numerical data \cite{Nishimoto2011,CritEntropy} although this is already possible knowing the general, though imprecise, nature of the corrections. That is, there is a well-defined peak in the central charge at BKT transitions as described and then implemented for periodic boundaries in \cite{JStats2010,Nishimoto2011} and then for open boundaries in \cite{CritEntropy}.
	
	In this article, we analytically derive the first order, finite-size correction to c at BKT transitions:
	
	\begin{equation}
		\label{nonlinearcurvefit}
		c_{\rm eff}(\ell) =1+R\left(\frac{y_0}{1+ y_0\ell}\right)^3
	\end{equation}
	
	\noindent in which $R = 3/16$ is a universal constant, and $y_0 = 1$ is the topological defect fugacity. For the remainder of the paper I use $\ell = log(L/\pi)$ at the midpoint of the spin chain. This universally modifies equation \ref{eq:SemiInfiniteEntropy} for BKT transitions by increasing the curvature of the entanglement. I follow this derivation with a careful numerical study of ground state entropy via exact diagonalization and DMRG for the isotropic XXZ model in one dimension with periodic boundary conditions.
	
	\textit{Analytical Derivation} I assume that, because they have to be consistent with Zamolodchikov's $c$-theorem, the leading log corrections to $c_{\rm eff}$ are the same independent of whether they measure the finite-size scaling on a cylinder, the entanglement of an interval, or any other physical quantity \cite{zamolodchikov1986ab,cardy1988conformal}.
	
	If I have a set of (almost) marginal couplings
	$\{g_k\}$ coupling to operators $\Phi_k$, the RG  equations in general take the form
	$$
	\dot g_k=y_kg_k+\pi\sum_{ij}c_{ijk}g_ig_j+\cdots
	$$
	where $y_k$ is the RG eigenvalue at the fixed point and $c_{ijk}$ is the coefficient in the OPE
	$$
	\Phi_i\cdot\Phi_j=\sum_kc_{ijk}\Phi_k \text{.}
	$$
	
	In the CFT normalization where the 2-point functions $\langle\Phi_i(r)\Phi_i(0)\rangle$ are normalized to 1 at separation $r=1$, the coefficients  $c_{ijk}$ are universal and symmetric in the indices.
	
	To this order the RG equations are then gradient flows
	$$
	\dot g_k=(\partial/\partial g_k)\widetilde C(\{g_i\})
	$$
	where 
	$$
	\widetilde C(\{g_i\})=\frac12\sum_ky_kg_k^2+\frac13\pi\sum_{ijk}c_{ijk}g_ig_jg_k+\cdots
	$$
	\color{red}
	
	\color{black}
	In terms of this Zamolodchikov's $c$-function is
	$$
	C(\{g_i\})=c-6\pi^2\widetilde C(\{g_i\})+\cdots \text{.}
	$$
	
	Now specialise to the BKT transition (or any model which maps onto it in the long-distance limit). There are two marginal operators. The CFT is defined by the gaussian model action
	$$
	S=\frac g{4\pi}\int(\partial\phi)^2d^2r
	\text{.}$$
	One marginal operator is $\propto(\partial\phi)^2$. The others are the vortex (anti-vortex) operators $V$ and $\bar V$
	with corresponding fugacity $y$. 
	
	The scaling dimension is $g/2$, so the RG equation for $y$ is
	$$
	\dot y=(2-g/2)y \text{.}
	$$
	The BKT point is $g=4$.
	
	We need to get the right normalization for $(\partial\phi)^2$. Using
	$$
	\langle\phi(r_1)\phi(r_2)\rangle=-(1/g)\log|r_1-r_2|+{\rm const.}
	$$
	and then from Wick's theorem, 
	$$
	\langle(\partial\phi(r_1))^2(\partial\phi(r_2))^2\rangle=\frac{4/g^2}{r^4} \text{.}
	$$
	
	Thus at the BKT point the correctly normalized operator  is $\Phi=2(\partial\phi)^2$. 
	
	The perturbation of the action is 
	$$
	\frac{\delta g}{4\pi}(\partial\phi)^2=x\Phi
	$$
	where $x=\delta g/(8\pi)$. In terms of this quantity the RG equation for $y$ becomes
	$$
	\dot y=-4\pi xy
	$$
	Thus 
	$$
	\widetilde C=-2\pi xy^2+\cdots
	$$
	and the other RG equation is therefore
	$$
	\dot x=-2\pi y^2
	$$
	Note that I have avoided working out the correct normalization for $V$. 
	
	The RG trajectory corresponding  to the BKT transition line is found by setting $y=\alpha x$, so, dividing the two equations we have 
	$\alpha=2/\alpha$, that is $\alpha=\sqrt2$.

	\begin{figure}
		\includegraphics[width=\linewidth]{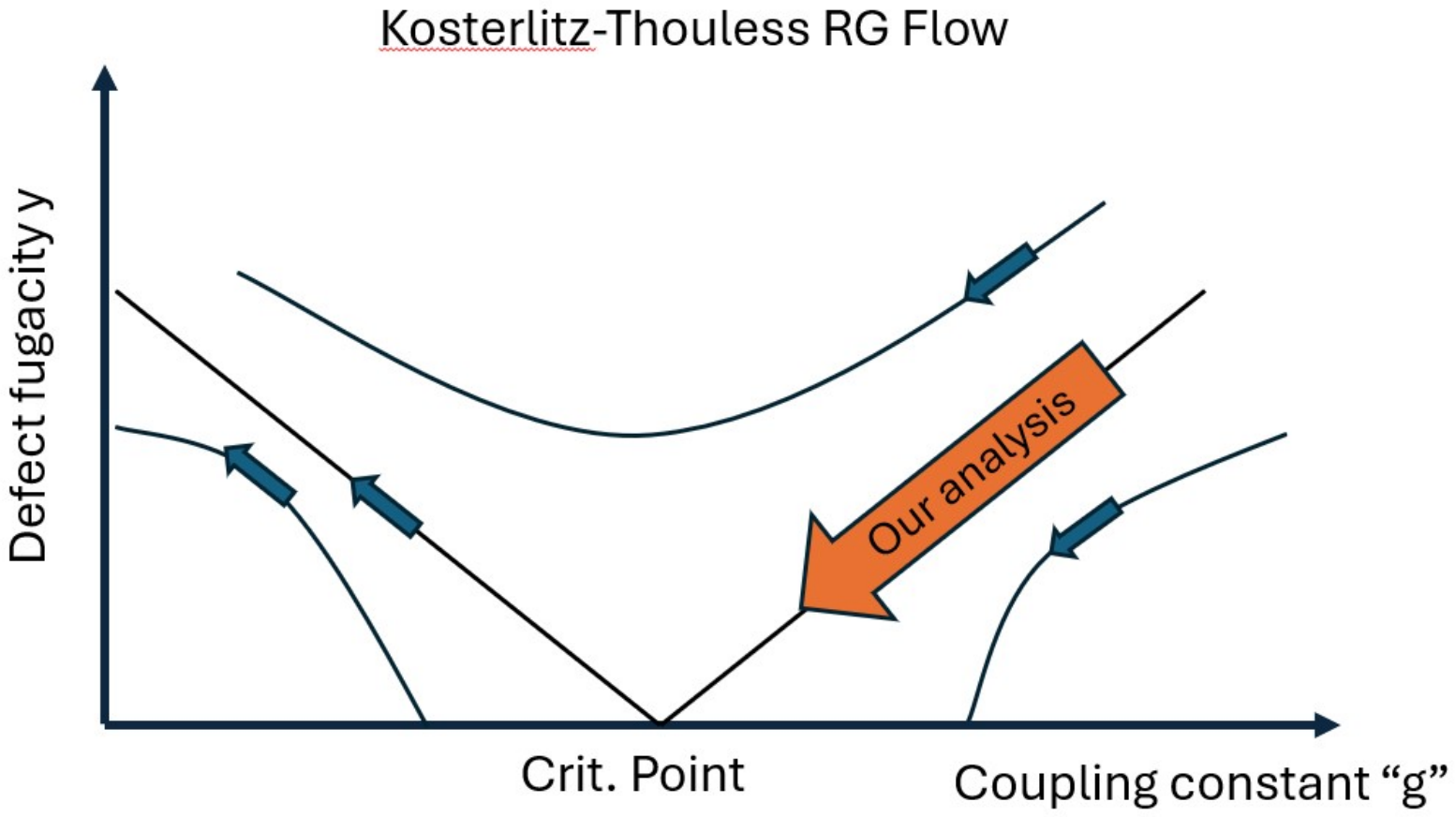}
		\caption{The RG flow is towards the BKT fixed point.}
		\label{RGflow}
	\end{figure}
	
	Along this trajectory, 
	$$
	\dot y=-2\sqrt 2\pi y^2
	$$
	so that
	$$
	y(\ell)=\frac{y_0}{1+2\sqrt 2\pi y_0\ell}
	$$
	where $\ell$ is the RG scale, $\sim \log L$ for the entanglement problem. 
	
	Along the transition line
	$$
	\widetilde C=-\sqrt 2\pi y^3
	$$
	so the c-function becomes
	$$
	c_{\rm eff}(\ell)=1+(6\pi^2)(\sqrt 2\pi)\left(\frac{y_0}{1+2\sqrt 2\pi y_0\ell}\right)^3
	$$
	Since $y_0$ (which is like the bare vortex fugacity) is arbitrary, we can rescale
	$2\sqrt 2\pi y_0\to y_0$
	in which case this becomes
	$$
	c_{\rm eff}(\ell)=1+\frac{3}{16}\left(\frac{y_0}{1+ y_0\ell}\right)^3
	$$
	
	To apply this to the entanglement problem for a single interval, I should recall that in the pure CFT
	
	$$
	S(L)\sim (c/3)\ell
	$$
	
	Here, I use the results from \cite{casini2004} that implies I may interpret that $\frac{\partial S}{ \partial \ell}$ obeys Zamolodchikov's c-theorem, that is, 
	
	$$c_{\rm eff}(\textit{l}) = 3\frac{\partial S}{ \partial \ell}\text{.}$$
	
	Then by integrating I get the next higher correction to the entanglement entropy,

	$$
	S(L)\sim (1/3)\log L-\frac{1}{16}\frac{y_0^2}{(1+y_0\log L)^2}+\cdots \text{.}
	$$
	
	Note that the value $R = 3/16$ might asymptotically grow to a larger value, though this is not clear from the relatively small system sizes I study next. 
	
	\textit{Numerical Study} In this section, I present numerical investigations that confirm the universal, analytical derivation in the prior section. Because the analytical corrections to c are universal at all finite-size BKT phase transitions, I choose a convenient model and method. That is, we extract c from entanglement entropy in the antiferromagnetic Heisenberg model in one spatial dimension with the DMRG method applied to periodic boundary conditions.
	
	The antiferromagnetic Heisenberg model can be thought of as the BKT transition separating a disordered critical phase from a gapped ordered phase at the isotropic point, $\Delta = 1$, in the spin-1/2 XXZ model:
	
	\begin{equation}
		H = \sum_{i} s^x_is^x_{i+1} + s^y_is^y_{i+1} + \Delta s^z_is^z_{i+1}\text{.}
	\end{equation}\label{eq:XXZmodel}
	
	The entanglement entropy is computed using the Schmidt decomposition of the ground-state wavefunction, measured between each pair of neighboring spins indexed by the spin at the left side. $S_j$ is then the entropy computed by cutting the chain between sites j and j+1 by summing over the Schmidt index i:
	
	\begin{equation}
		S_{j} = -\sum_{i} \lambda_{i}^2 \log{(\lambda_{i}^2)}\text{.}
	\end{equation}\label{eq:VNEntanglement}
	
	This is very convenient in DMRG because the algorithm uses the Schmidt values for performing the truncation scheme, and so the Schmidt values are readily available. From this data, I evaluate the derivative with respect to j at the midpoint j = L/2. Note that there are subtleties of relating the discrete index $j$ to the continuum variable $x$, however this is negligible when $x = L/2$.
	
	Once I have central charge for each system size I performed both nonlinear and linear regressions to fit to the analysis derived above. The curve fitting parameters include the bare topological defect fugacity, $y_0$, and a universal coefficient that I arbitrarily call ``R":
	
	\begin{equation}
		\label{eq:nonlinearcurvefit}
		c_{\rm eff}(\ell) =1+R\left(\frac{y_0}{1+ y_0\ell}\right)^3\text{.}
	\end{equation}
	
	As mentioned, this study is restricted to periodic boundary conditions which are known to drastically reduce the accessible system sizes for the DMRG algorithm. As a result, I have reliable data for systems up to length 120 sites. For sizes under 32 sites, I used exact diagonalization to provide a check against the convergence accuracy of the DMRG results, as shown in the appendix.
	
\begin{figure}[t!]\centering
		\includegraphics[width=\linewidth]{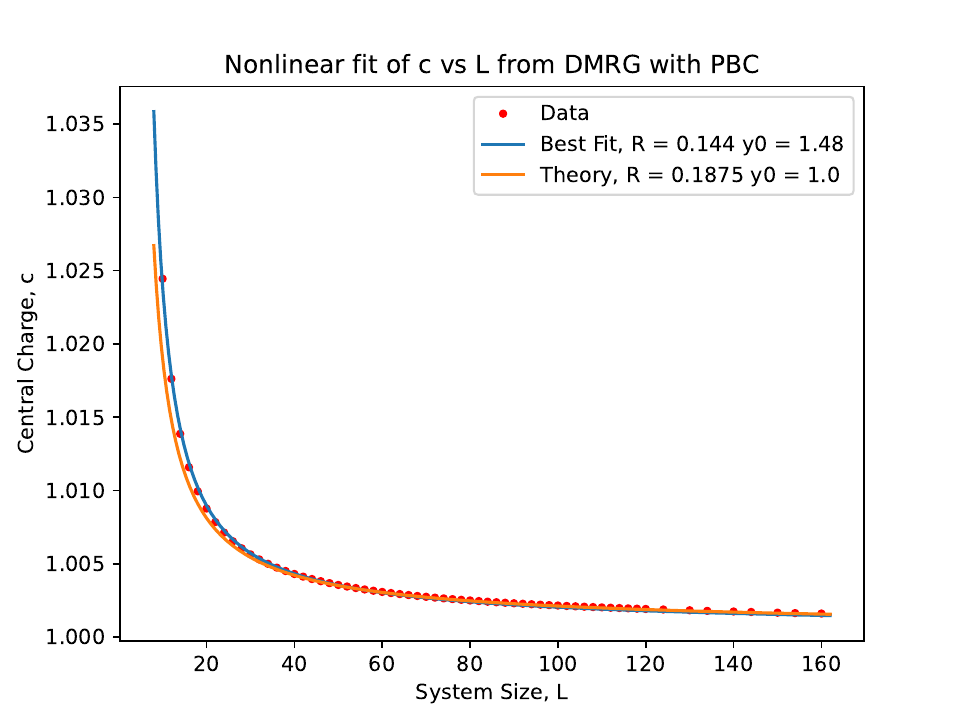}
		\caption{Curve fit of $c_{eff}$ in equation (\ref{eq:nonlinearcurvefit}) at chain midpoint for all sizes available.}
		\label{fig:plain_c}
	\end{figure}

	The first data I present is a plot of the central charge vs. system size, along with a nonlinear curve fit to extract the two fitting parameters of interest (fig. \ref{fig:plain_c}).
	
	One of the conclusions drawn from that plot is that plotting and fitting functions with logarithmic dependencies can be tricky, so I also found a linearized version by transforming the variables as follows. Let $Y \equiv \frac{1}{y_0}$, $M \equiv \frac{1}{R^{1/3}}$, and $\ell \equiv \log\left(\frac{L}{\pi}\sin{(\pi x/L)}\right) = \log{(L/\pi)}$. Then the fitting function becomes a much simpler
	
	\begin{equation}\label{eq:trans_c}
		C = M(Y + \ell)
	\end{equation}
	
	with dependent variable
	
	\begin{equation}
		C \equiv \frac{1}{(c(x,L)-1)^{1/3}}\text{.}
	\end{equation}\label{eq:linearcurvefit}
	
\begin{figure}[t!]\centering
		\includegraphics[width=\linewidth]{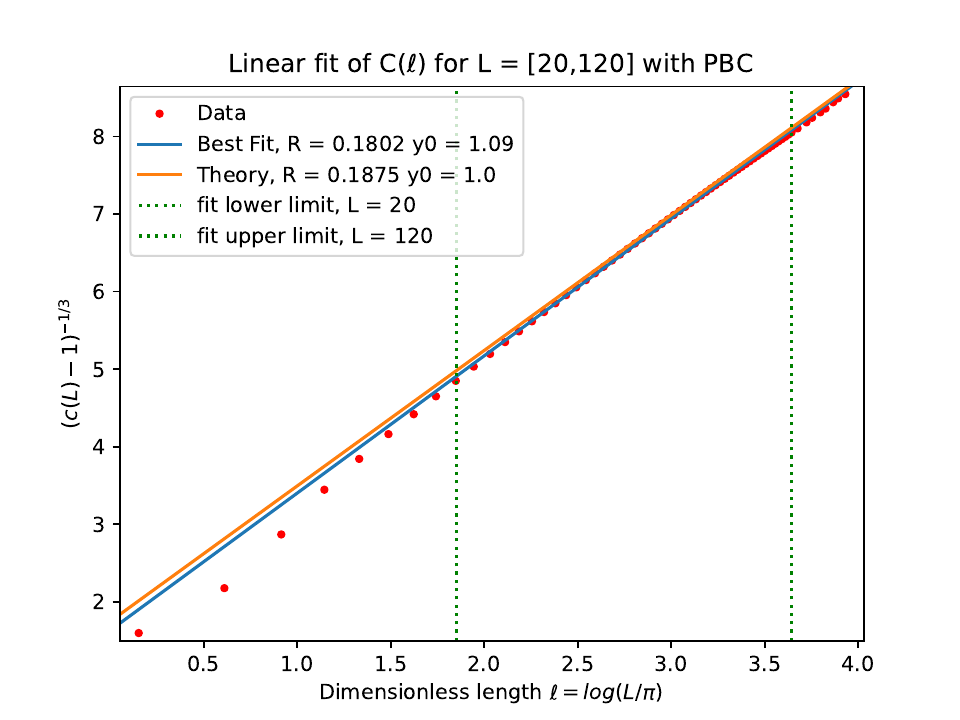}
		\caption{Linear fit of $c_{eff}$ transformed to the linear form, equation (\ref{eq:trans_c}). }
		\label{fig:linear_c}
	\end{figure}

With this form, figure \ref{fig:plain_c} becomes figure \ref{fig:linear_c} from which it is clear that equation \ref{nonlinearcurvefit} is the correct fitting function to match the data, including a nonzero intercept on the $C$ axis obtained at the non-physical $\ell = 0$.
	
	In the appendix, I provide additional investigations of these linear regressions to reinforce the conclusions drawn in the paper. 
	
	\textit{Conclusions} In this paper I have provided an elegant derivation that defines the central charge as the first derivative of entropy with respect to logarithm of length scale. As such, this ``effective" finite-size corrected value of c will apply equally to classical and quantum critical systems with a two-dimensional conformal field description. Also, I have only considered corrections to c due to finite size effects, but other parameters (fields, temperature, truncation error in finite and infinite MPS) should also follow.
	
	The case of spin chains with open boundaries provides another opportunity to analyze finite-size corrections to central charge as I have done here for periodic boundaries. My prior work \cite{CritEntropy} provides a starting point to these studies, including methods to deal with bond-alternation in the entropy as a function of position as well as strategies to extrapolate curve fits to the middle of the lattice, named ``scaling to the middle." Note that equation \ref{eq:SemiInfiniteEntropy} is modified for open boundaries by replacing the 3 with a 6 and multiplying L by 2 within the logarithm \cite{CritEntropy}.
	
\begin{figure}[h]\centering
		\includegraphics[width=\linewidth]{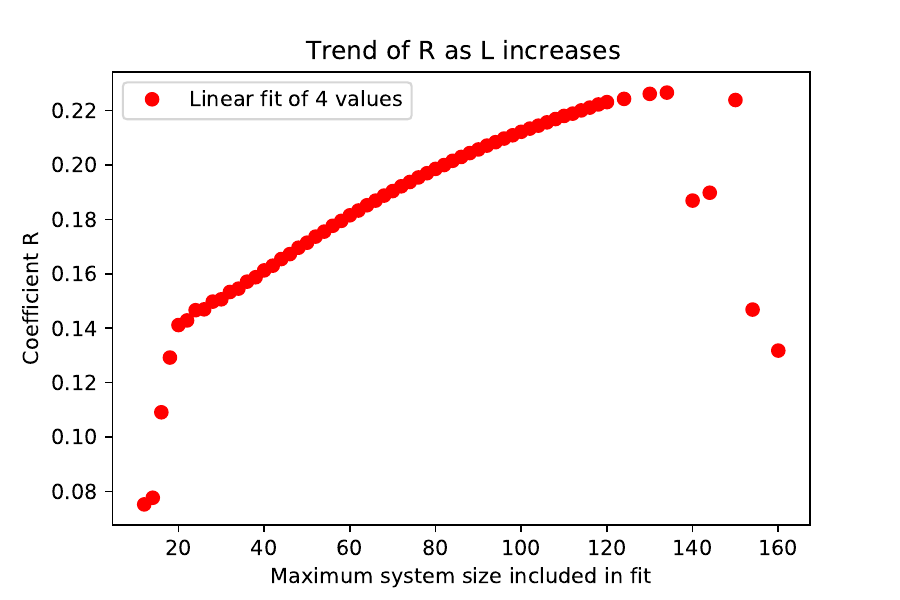}
		\caption{Plot of the curve fit parameter, R, for each possible set of 4 neighboring data points as a function of system size. R is clearly increasing as system size increases. Fugacity $y_0$ is allowed to vary. From this perspective it appears R might asymptotically increase to the value 3/8.}
		\label{fig:RvsL}
	\end{figure}

\begin{figure}[h]\centering
		\includegraphics[width=\linewidth]{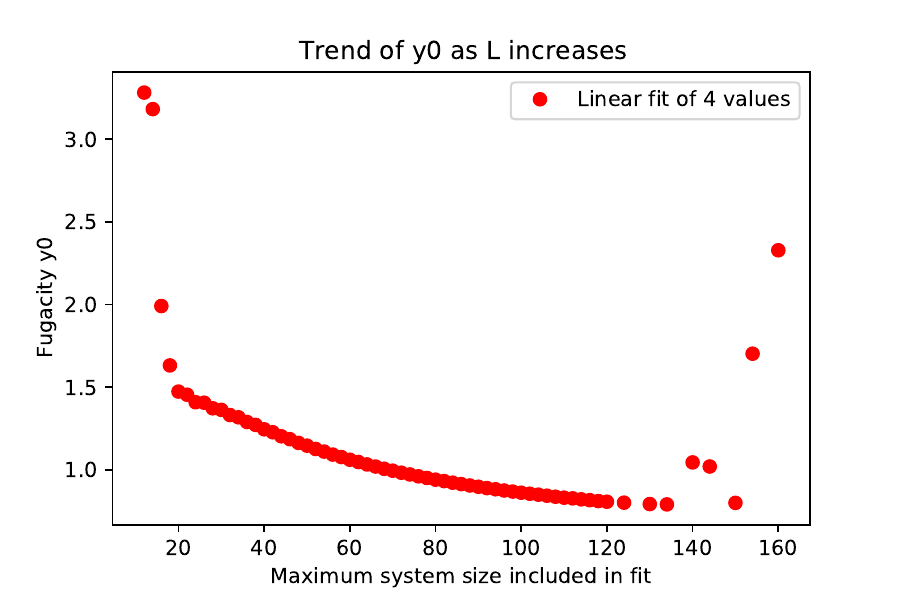}
		\caption{$y_0$ plotted as a function of system size. Like for coefficient R plotted above, $y_0$ has poor results for sizes less than 20 and greater than 120, so in this plot we restrict to the smooth portion of the data.}
		\label{fig:y0vsL}
	\end{figure}

	Although the results presented here offer reliable measurements for central charge for the system sizes studied, it is possible that the asymptotic value of $c_{eff}$ for larger sizes may vary as indicated by figures \ref{fig:RvsL} and \ref{fig:y0vsL} when both R and $y_0$ are allowed to vary. If these trends hold for larger size spin chains, additional terms in the operator product expansion may be needed to account for higher-order finite-size effects. 
	
	In addition, small system sizes ($L < 30$) may have analytically-tractable contributions from the operator product expansion that provide additional insight into the nature of topological defects.
	
	Applying this approach to other defect-driven phase transitions is yet another possible extension of this work.
	
	\textit{Acknowledgements} I thank John Cardy for the renormalization group analysis presented here for BKT transitions. This research was supported in part by the NSF under grant DMR-1411345 and by University of California, Riverside's GRMP fellowship. This work used the Extreme Science and Engineering Discovery Environment (XSEDE) COMET at the San Diego Supercomputer Center through allocation TG-DMR170082 \cite{EXCEDEallocation} as well as University of California Riverside's High Perfomance Computing Center.
	
	\textit{Appendix}
	
	\textit{Study of size-dependence of curve fit parameters}

	\begin{figure}[t!]\centering
		\includegraphics[width=\linewidth]{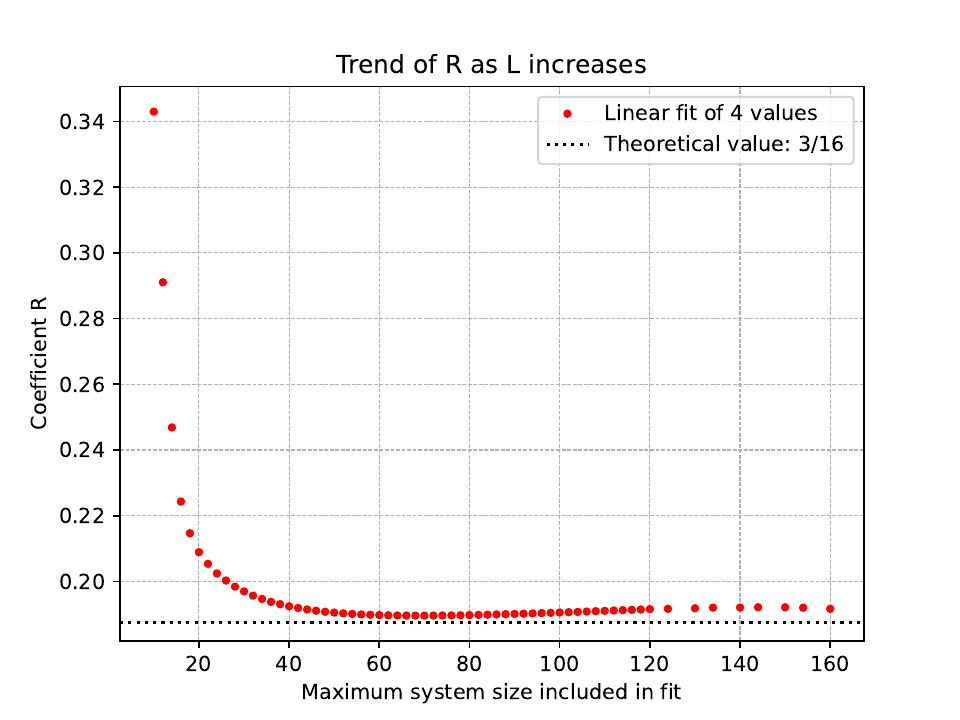}
		\caption{Plot of the curve fit parameter, R, for each possible set of 4 neighboring data points as a function of system size, including very small and very large. Fugacity $y_0$ is fixed to 1.}
		\label{fig:RvsLy0fixed}
	\end{figure}
	
	\FloatBarrier
	
	\textit{Study of Truncation Effects in Curve Fit Results}

	\begin{figure}[h] 
		\includegraphics[width=\linewidth]{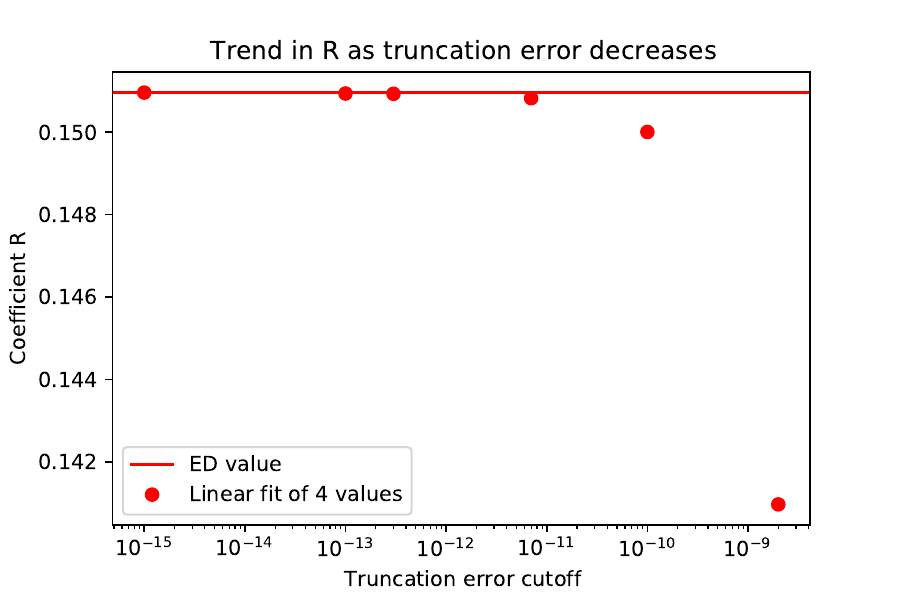}
		\caption{Plot of coefficient R vs the DMRG truncation parameter for system sizes 24 to 30 plotted with the exact diagonalization result. The error in R is at the $5^{th}$ decimal place.}
		\label{fig:RvsE30}
	\end{figure}

	\begin{figure}  
		\includegraphics[width=\linewidth]{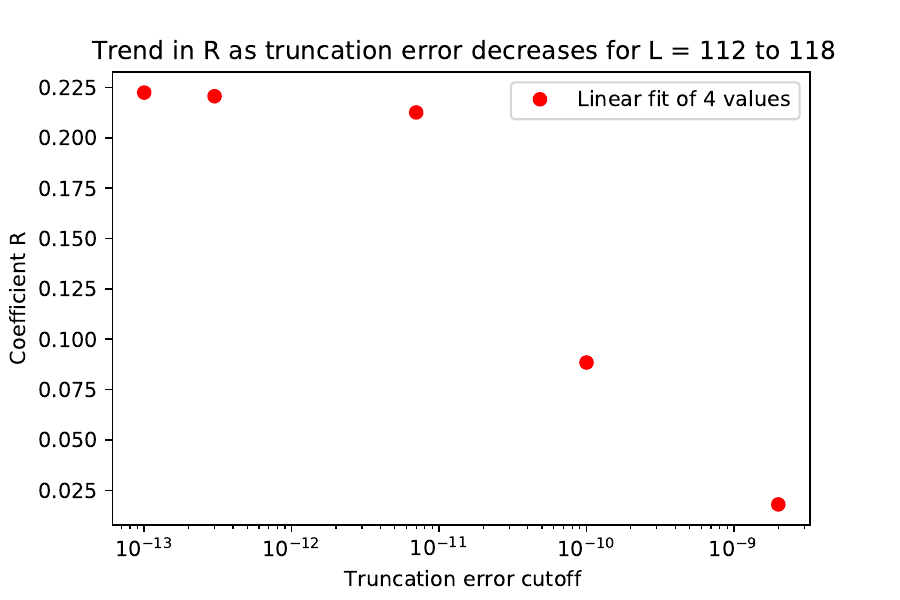}
		\caption{Similar to figure \ref{fig:RvsE30} but for system sizes 112 to 118. The error in R for our data is approximately the difference between the first two data points in this figure, which is less than 0.001, but still discernible in this plot. It is noteworthy that the same truncation error in DMRG results in a larger error in R when the system size is increased; error increases by factor of 100.}
		\label{fig:RvsE120}
	\end{figure}
	
	\FloatBarrier
	\bibliographystyle{apsrev4-2}
	\bibliography{c_eff.bib,bkt.bib}

\begin{thebibliography}{26}%
\makeatletter
\providecommand \@ifxundefined [1]{%
 \@ifx{#1\undefined}
}%
\providecommand \@ifnum [1]{%
 \ifnum #1\expandafter \@firstoftwo
 \else \expandafter \@secondoftwo
 \fi
}%
\providecommand \@ifx [1]{%
 \ifx #1\expandafter \@firstoftwo
 \else \expandafter \@secondoftwo
 \fi
}%
\providecommand \natexlab [1]{#1}%
\providecommand \enquote  [1]{``#1''}%
\providecommand \bibnamefont  [1]{#1}%
\providecommand \bibfnamefont [1]{#1}%
\providecommand \citenamefont [1]{#1}%
\providecommand \href@noop [0]{\@secondoftwo}%
\providecommand \href [0]{\begingroup \@sanitize@url \@href}%
\providecommand \@href[1]{\@@startlink{#1}\@@href}%
\providecommand \@@href[1]{\endgroup#1\@@endlink}%
\providecommand \@sanitize@url [0]{\catcode `\\12\catcode `\$12\catcode
  `\&12\catcode `\#12\catcode `\^12\catcode `\_12\catcode `\%12\relax}%
\providecommand \@@startlink[1]{}%
\providecommand \@@endlink[0]{}%
\providecommand \url  [0]{\begingroup\@sanitize@url \@url }%
\providecommand \@url [1]{\endgroup\@href {#1}{\urlprefix }}%
\providecommand \urlprefix  [0]{URL }%
\providecommand \Eprint [0]{\href }%
\providecommand \doibase [0]{https://doi.org/}%
\providecommand \selectlanguage [0]{\@gobble}%
\providecommand \bibinfo  [0]{\@secondoftwo}%
\providecommand \bibfield  [0]{\@secondoftwo}%
\providecommand \translation [1]{[#1]}%
\providecommand \BibitemOpen [0]{}%
\providecommand \bibitemStop [0]{}%
\providecommand \bibitemNoStop [0]{.\EOS\space}%
\providecommand \EOS [0]{\spacefactor3000\relax}%
\providecommand \BibitemShut  [1]{\csname bibitem#1\endcsname}%
\let\auto@bib@innerbib\@empty
\bibitem [{\citenamefont {Francesco}\ \emph {et~al.}(1997)\citenamefont
  {Francesco}, \citenamefont {Mathieu},\ and\ \citenamefont
  {S{\'e}n{\'e}chal}}]{FMSCFT}%
  \BibitemOpen
  \bibfield  {author} {\bibinfo {author} {\bibfnamefont {P.~D.}\ \bibnamefont
  {Francesco}}, \bibinfo {author} {\bibfnamefont {P.}~\bibnamefont {Mathieu}},\
  and\ \bibinfo {author} {\bibfnamefont {D.}~\bibnamefont {S{\'e}n{\'e}chal}},\
  }\href@noop {} {\emph {\bibinfo {title} {Conformal Field Theory}}}\ (\bibinfo
   {publisher} {Springer},\ \bibinfo {year} {1997})\BibitemShut {NoStop}%
\bibitem [{\citenamefont {Bl\"ote}\ \emph {et~al.}(1986)\citenamefont
  {Bl\"ote}, \citenamefont {Cardy},\ and\ \citenamefont
  {Nightingale}}]{PhysRevLett.56.742}%
  \BibitemOpen
  \bibfield  {author} {\bibinfo {author} {\bibfnamefont {H.~W.~J.}\
  \bibnamefont {Bl\"ote}}, \bibinfo {author} {\bibfnamefont {J.~L.}\
  \bibnamefont {Cardy}},\ and\ \bibinfo {author} {\bibfnamefont {M.~P.}\
  \bibnamefont {Nightingale}},\ }\href@noop {} {\bibfield  {journal} {\bibinfo
  {journal} {Phys. Rev. Lett.}\ }\textbf {\bibinfo {volume} {56}},\ \bibinfo
  {pages} {742} (\bibinfo {year} {1986})}\BibitemShut {NoStop}%
\bibitem [{\citenamefont {Affleck}(1986)}]{PhysRevLett.56.746}%
  \BibitemOpen
  \bibfield  {author} {\bibinfo {author} {\bibfnamefont {I.}~\bibnamefont
  {Affleck}},\ }\href@noop {} {\bibfield  {journal} {\bibinfo  {journal} {Phys.
  Rev. Lett.}\ }\textbf {\bibinfo {volume} {56}},\ \bibinfo {pages} {746}
  (\bibinfo {year} {1986})}\BibitemShut {NoStop}%
\bibitem [{\citenamefont {Holzhey}\ \emph {et~al.}(1994)\citenamefont
  {Holzhey}, \citenamefont {Larsen},\ and\ \citenamefont
  {Wilczek}}]{HOLZHEY1994443}%
  \BibitemOpen
  \bibfield  {author} {\bibinfo {author} {\bibfnamefont {C.}~\bibnamefont
  {Holzhey}}, \bibinfo {author} {\bibfnamefont {F.}~\bibnamefont {Larsen}},\
  and\ \bibinfo {author} {\bibfnamefont {F.}~\bibnamefont {Wilczek}},\
  }\href@noop {} {\bibfield  {journal} {\bibinfo  {journal} {Nuclear Physics
  B}\ }\textbf {\bibinfo {volume} {424}},\ \bibinfo {pages} {443} (\bibinfo
  {year} {1994})}\BibitemShut {NoStop}%
\bibitem [{\citenamefont {Vidal}\ \emph {et~al.}(2003)\citenamefont {Vidal},
  \citenamefont {Latorre}, \citenamefont {Rico},\ and\ \citenamefont
  {Kitaev}}]{PhysRevLett.90.227902}%
  \BibitemOpen
  \bibfield  {author} {\bibinfo {author} {\bibfnamefont {G.}~\bibnamefont
  {Vidal}}, \bibinfo {author} {\bibfnamefont {J.~I.}\ \bibnamefont {Latorre}},
  \bibinfo {author} {\bibfnamefont {E.}~\bibnamefont {Rico}},\ and\ \bibinfo
  {author} {\bibfnamefont {A.}~\bibnamefont {Kitaev}},\ }\href@noop {}
  {\bibfield  {journal} {\bibinfo  {journal} {Phys. Rev. Lett.}\ }\textbf
  {\bibinfo {volume} {90}},\ \bibinfo {pages} {227902} (\bibinfo {year}
  {2003})}\BibitemShut {NoStop}%
\bibitem [{\citenamefont
  {Korepin}(2004{\natexlab{a}})}]{PhysRevLett.92.096402}%
  \BibitemOpen
  \bibfield  {author} {\bibinfo {author} {\bibfnamefont {V.~E.}\ \bibnamefont
  {Korepin}},\ }\href@noop {} {\bibfield  {journal} {\bibinfo  {journal} {Phys.
  Rev. Lett.}\ }\textbf {\bibinfo {volume} {92}},\ \bibinfo {pages} {096402}
  (\bibinfo {year} {2004}{\natexlab{a}})}\BibitemShut {NoStop}%
\bibitem [{\citenamefont {Jin}\ and\ \citenamefont
  {Korepin}(2004)}]{jin2004quantum}%
  \BibitemOpen
  \bibfield  {author} {\bibinfo {author} {\bibfnamefont {B.-Q.}\ \bibnamefont
  {Jin}}\ and\ \bibinfo {author} {\bibfnamefont {V.~E.}\ \bibnamefont
  {Korepin}},\ }\href@noop {} {\bibfield  {journal} {\bibinfo  {journal}
  {Journal of statistical physics}\ }\textbf {\bibinfo {volume} {116}},\
  \bibinfo {pages} {79} (\bibinfo {year} {2004})}\BibitemShut {NoStop}%
\bibitem [{\citenamefont
  {Korepin}(2004{\natexlab{b}})}]{korepin2004universality}%
  \BibitemOpen
  \bibfield  {author} {\bibinfo {author} {\bibfnamefont {V.~E.}\ \bibnamefont
  {Korepin}},\ }\href@noop {} {\bibfield  {journal} {\bibinfo  {journal}
  {Physical review letters}\ }\textbf {\bibinfo {volume} {92}},\ \bibinfo
  {pages} {096402} (\bibinfo {year} {2004}{\natexlab{b}})}\BibitemShut
  {NoStop}%
\bibitem [{\citenamefont {White}(1992)}]{white1992density}%
  \BibitemOpen
  \bibfield  {author} {\bibinfo {author} {\bibfnamefont {S.~R.}\ \bibnamefont
  {White}},\ }\href@noop {} {\bibfield  {journal} {\bibinfo  {journal}
  {Physical review letters}\ }\textbf {\bibinfo {volume} {69}},\ \bibinfo
  {pages} {2863} (\bibinfo {year} {1992})}\BibitemShut {NoStop}%
\bibitem [{\citenamefont {White}(1993)}]{white1993density}%
  \BibitemOpen
  \bibfield  {author} {\bibinfo {author} {\bibfnamefont {S.~R.}\ \bibnamefont
  {White}},\ }\href@noop {} {\bibfield  {journal} {\bibinfo  {journal}
  {Physical Review B}\ }\textbf {\bibinfo {volume} {48}},\ \bibinfo {pages}
  {10345} (\bibinfo {year} {1993})}\BibitemShut {NoStop}%
\bibitem [{\citenamefont {Schollwöck}(2011)}]{SCHOLLWOCK2011}%
  \BibitemOpen
  \bibfield  {author} {\bibinfo {author} {\bibfnamefont {U.}~\bibnamefont
  {Schollwöck}},\ }\href
  {https://doi.org/https://doi.org/10.1016/j.aop.2010.09.012} {\bibfield
  {journal} {\bibinfo  {journal} {Annals of Physics}\ }\textbf {\bibinfo
  {volume} {326}},\ \bibinfo {pages} {96 } (\bibinfo {year} {2011})},\ \bibinfo
  {note} {january 2011 Special Issue}\BibitemShut {NoStop}%
\bibitem [{\citenamefont {Hastings}\ \emph {et~al.}(2010)\citenamefont
  {Hastings}, \citenamefont {Gonz{\'a}lez}, \citenamefont {Kallin},\ and\
  \citenamefont {Melko}}]{hastings2010measuring}%
  \BibitemOpen
  \bibfield  {author} {\bibinfo {author} {\bibfnamefont {M.~B.}\ \bibnamefont
  {Hastings}}, \bibinfo {author} {\bibfnamefont {I.}~\bibnamefont
  {Gonz{\'a}lez}}, \bibinfo {author} {\bibfnamefont {A.~B.}\ \bibnamefont
  {Kallin}},\ and\ \bibinfo {author} {\bibfnamefont {R.~G.}\ \bibnamefont
  {Melko}},\ }\href@noop {} {\bibfield  {journal} {\bibinfo  {journal}
  {Physical review letters}\ }\textbf {\bibinfo {volume} {104}},\ \bibinfo
  {pages} {157201} (\bibinfo {year} {2010})}\BibitemShut {NoStop}%
\bibitem [{\citenamefont {Kosterlitz}\ and\ \citenamefont
  {Thouless}(2018)}]{kosterlitz2018ordering}%
  \BibitemOpen
  \bibfield  {author} {\bibinfo {author} {\bibfnamefont {J.~M.}\ \bibnamefont
  {Kosterlitz}}\ and\ \bibinfo {author} {\bibfnamefont {D.~J.}\ \bibnamefont
  {Thouless}},\ }in\ \href@noop {} {\emph {\bibinfo {booktitle} {Basic Notions
  Of Condensed Matter Physics}}}\ (\bibinfo  {publisher} {CRC Press},\ \bibinfo
  {year} {2018})\ pp.\ \bibinfo {pages} {493--515}\BibitemShut {NoStop}%
\bibitem [{\citenamefont {Berezinskii}(1971)}]{berezinskii1971destruction}%
  \BibitemOpen
  \bibfield  {author} {\bibinfo {author} {\bibfnamefont {V.}~\bibnamefont
  {Berezinskii}},\ }\href@noop {} {\bibfield  {journal} {\bibinfo  {journal}
  {Sov. Phys. JETP}\ }\textbf {\bibinfo {volume} {32}},\ \bibinfo {pages} {493}
  (\bibinfo {year} {1971})}\BibitemShut {NoStop}%
\bibitem [{\citenamefont {Berezinskii}(1972)}]{berezinskii1972destruction}%
  \BibitemOpen
  \bibfield  {author} {\bibinfo {author} {\bibfnamefont {V.}~\bibnamefont
  {Berezinskii}},\ }\href@noop {} {\bibfield  {journal} {\bibinfo  {journal}
  {Sov. Phys. JETP}\ }\textbf {\bibinfo {volume} {34}},\ \bibinfo {pages} {610}
  (\bibinfo {year} {1972})}\BibitemShut {NoStop}%
\bibitem [{\citenamefont {Laflorencie}\ \emph {et~al.}(2006)\citenamefont
  {Laflorencie}, \citenamefont {S\o{}rensen}, \citenamefont {Chang},\ and\
  \citenamefont {Affleck}}]{Laflorencie2006}%
  \BibitemOpen
  \bibfield  {author} {\bibinfo {author} {\bibfnamefont {N.}~\bibnamefont
  {Laflorencie}}, \bibinfo {author} {\bibfnamefont {E.~S.}\ \bibnamefont
  {S\o{}rensen}}, \bibinfo {author} {\bibfnamefont {M.-S.}\ \bibnamefont
  {Chang}},\ and\ \bibinfo {author} {\bibfnamefont {I.}~\bibnamefont
  {Affleck}},\ }\href {https://doi.org/10.1103/PhysRevLett.96.100603}
  {\bibfield  {journal} {\bibinfo  {journal} {Phys. Rev. Lett.}\ }\textbf
  {\bibinfo {volume} {96}},\ \bibinfo {pages} {100603} (\bibinfo {year}
  {2006})}\BibitemShut {NoStop}%
\bibitem [{\citenamefont {Cardy}\ and\ \citenamefont
  {Calabrese}(2010)}]{JStats2010}%
  \BibitemOpen
  \bibfield  {author} {\bibinfo {author} {\bibfnamefont {J.}~\bibnamefont
  {Cardy}}\ and\ \bibinfo {author} {\bibfnamefont {P.}~\bibnamefont
  {Calabrese}},\ }\href {http://stacks.iop.org/1742-5468/2010/i=04/a=P04023}
  {\bibfield  {journal} {\bibinfo  {journal} {Journal of Statistical Mechanics:
  Theory and Experiment}\ }\textbf {\bibinfo {volume} {2010}},\ \bibinfo
  {pages} {P04023} (\bibinfo {year} {2010})}\BibitemShut {NoStop}%
\bibitem [{\citenamefont {Fagotti}\ and\ \citenamefont
  {Calabrese}(2011)}]{JStats2011P01017}%
  \BibitemOpen
  \bibfield  {author} {\bibinfo {author} {\bibfnamefont {M.}~\bibnamefont
  {Fagotti}}\ and\ \bibinfo {author} {\bibfnamefont {P.}~\bibnamefont
  {Calabrese}},\ }\href {http://stacks.iop.org/1742-5468/2011/i=01/a=P01017}
  {\bibfield  {journal} {\bibinfo  {journal} {Journal of Statistical Mechanics:
  Theory and Experiment}\ }\textbf {\bibinfo {volume} {2011}},\ \bibinfo
  {pages} {P01017} (\bibinfo {year} {2011})}\BibitemShut {NoStop}%
\bibitem [{\citenamefont {Calabrese}\ and\ \citenamefont
  {Cardy}(2009)}]{calabrese2009entanglement}%
  \BibitemOpen
  \bibfield  {author} {\bibinfo {author} {\bibfnamefont {P.}~\bibnamefont
  {Calabrese}}\ and\ \bibinfo {author} {\bibfnamefont {J.}~\bibnamefont
  {Cardy}},\ }\href@noop {} {\bibfield  {journal} {\bibinfo  {journal} {Journal
  of Physics A: Mathematical and Theoretical}\ }\textbf {\bibinfo {volume}
  {42}},\ \bibinfo {pages} {504005} (\bibinfo {year} {2009})}\BibitemShut
  {NoStop}%
\bibitem [{\citenamefont {Casini}\ and\ \citenamefont
  {Huerta}(2004)}]{casini2004}%
  \BibitemOpen
  \bibfield  {author} {\bibinfo {author} {\bibfnamefont {H.}~\bibnamefont
  {Casini}}\ and\ \bibinfo {author} {\bibfnamefont {M.}~\bibnamefont
  {Huerta}},\ }\href
  {https://doi.org/https://doi.org/10.1016/j.physletb.2004.08.072} {\bibfield
  {journal} {\bibinfo  {journal} {Physics Letters B}\ }\textbf {\bibinfo
  {volume} {600}},\ \bibinfo {pages} {142 } (\bibinfo {year}
  {2004})}\BibitemShut {NoStop}%
\bibitem [{\citenamefont {Calabrese}\ \emph {et~al.}(2010)\citenamefont
  {Calabrese}, \citenamefont {Campostrini}, \citenamefont {Essler},\ and\
  \citenamefont {Nienhuis}}]{PhysRevLett.104.095701}%
  \BibitemOpen
  \bibfield  {author} {\bibinfo {author} {\bibfnamefont {P.}~\bibnamefont
  {Calabrese}}, \bibinfo {author} {\bibfnamefont {M.}~\bibnamefont
  {Campostrini}}, \bibinfo {author} {\bibfnamefont {F.}~\bibnamefont
  {Essler}},\ and\ \bibinfo {author} {\bibfnamefont {B.}~\bibnamefont
  {Nienhuis}},\ }\href {https://doi.org/10.1103/PhysRevLett.104.095701}
  {\bibfield  {journal} {\bibinfo  {journal} {Phys. Rev. Lett.}\ }\textbf
  {\bibinfo {volume} {104}},\ \bibinfo {pages} {095701} (\bibinfo {year}
  {2010})}\BibitemShut {NoStop}%
\bibitem [{\citenamefont {Nishimoto}(2011)}]{Nishimoto2011}%
  \BibitemOpen
  \bibfield  {author} {\bibinfo {author} {\bibfnamefont {S.}~\bibnamefont
  {Nishimoto}},\ }\href {https://doi.org/10.1103/PhysRevB.84.195108} {\bibfield
   {journal} {\bibinfo  {journal} {Phys. Rev. B}\ }\textbf {\bibinfo {volume}
  {84}},\ \bibinfo {pages} {195108} (\bibinfo {year} {2011})}\BibitemShut
  {NoStop}%
\bibitem [{\citenamefont {Spalding}\ \emph {et~al.}(2019)\citenamefont
  {Spalding}, \citenamefont {Tsai},\ and\ \citenamefont
  {Campbell}}]{CritEntropy}%
  \BibitemOpen
  \bibfield  {author} {\bibinfo {author} {\bibfnamefont {J.}~\bibnamefont
  {Spalding}}, \bibinfo {author} {\bibfnamefont {S.-W.}\ \bibnamefont {Tsai}},\
  and\ \bibinfo {author} {\bibfnamefont {D.~K.}\ \bibnamefont {Campbell}},\
  }\href {https://doi.org/10.1103/PhysRevB.99.195445} {\bibfield  {journal}
  {\bibinfo  {journal} {Phys. Rev. B}\ }\textbf {\bibinfo {volume} {99}},\
  \bibinfo {pages} {195445} (\bibinfo {year} {2019})}\BibitemShut {NoStop}%
\bibitem [{\citenamefont {Zamolodchikov}(1986)}]{zamolodchikov1986ab}%
  \BibitemOpen
  \bibfield  {author} {\bibinfo {author} {\bibfnamefont {A.~B.}\ \bibnamefont
  {Zamolodchikov}},\ }\href@noop {} {\bibfield  {journal} {\bibinfo  {journal}
  {JETP Lett.}\ }\textbf {\bibinfo {volume} {43}},\ \bibinfo {pages} {731}
  (\bibinfo {year} {1986})}\BibitemShut {NoStop}%
\bibitem [{\citenamefont {Cardy}(1988)}]{cardy1988conformal}%
  \BibitemOpen
  \bibfield  {author} {\bibinfo {author} {\bibfnamefont {J.~L.}\ \bibnamefont
  {Cardy}},\ }\href@noop {} {\bibfield  {journal} {\bibinfo  {journal} {Les
  Houches}\ } (\bibinfo {year} {1988})}\BibitemShut {NoStop}%
\bibitem [{\citenamefont {Towns}\ \emph {et~al.}(2014)\citenamefont {Towns},
  \citenamefont {Cockerill}, \citenamefont {Dahan}, \citenamefont {Foster},
  \citenamefont {Gaither}, \citenamefont {Grimshaw}, \citenamefont {Hazlewood},
  \citenamefont {Lathrop}, \citenamefont {Lifka}, \citenamefont {Peterson},
  \citenamefont {Roskies}, \citenamefont {Scott},\ and\ \citenamefont
  {Wilkins-Diehr}}]{EXCEDEallocation}%
  \BibitemOpen
  \bibfield  {author} {\bibinfo {author} {\bibfnamefont {J.}~\bibnamefont
  {Towns}}, \bibinfo {author} {\bibfnamefont {T.}~\bibnamefont {Cockerill}},
  \bibinfo {author} {\bibfnamefont {M.}~\bibnamefont {Dahan}}, \bibinfo
  {author} {\bibfnamefont {I.}~\bibnamefont {Foster}}, \bibinfo {author}
  {\bibfnamefont {K.}~\bibnamefont {Gaither}}, \bibinfo {author} {\bibfnamefont
  {A.}~\bibnamefont {Grimshaw}}, \bibinfo {author} {\bibfnamefont
  {V.}~\bibnamefont {Hazlewood}}, \bibinfo {author} {\bibfnamefont
  {S.}~\bibnamefont {Lathrop}}, \bibinfo {author} {\bibfnamefont
  {D.}~\bibnamefont {Lifka}}, \bibinfo {author} {\bibfnamefont {G.~D.}\
  \bibnamefont {Peterson}}, \bibinfo {author} {\bibfnamefont {R.}~\bibnamefont
  {Roskies}}, \bibinfo {author} {\bibfnamefont {J.~R.}\ \bibnamefont {Scott}},\
  and\ \bibinfo {author} {\bibfnamefont {N.}~\bibnamefont {Wilkins-Diehr}},\
  }\href@noop {} {\bibfield  {journal} {\bibinfo  {journal} {Computing in
  Science \& Engineering}\ }\textbf {\bibinfo {volume} {16}},\ \bibinfo {pages}
  {62} (\bibinfo {year} {2014})}\BibitemShut {NoStop}%
\end{thebibliography}%
	
\end{document}